\documentclass[twocolumn,showpacs,superscriptaddress,preprintnumbers,amsmath,amssymb,prc]{revtex4-1}

\usepackage{graphicx}
\usepackage{dcolumn}
\usepackage{bm}
\usepackage{ifpdf}
\usepackage{epstopdf}
\usepackage{slashed}
\usepackage{amsfonts}
\usepackage{mathrsfs}
\usepackage{amssymb}
\usepackage{multirow}
\usepackage{CJK}
\usepackage{xcolor}

\begin{document}

\preprint{RIKEN-QHP-436}
\preprint{RIKEN-iTHEMS-Report-20}

\title{Non-relativistic expansion of Dirac equation with spherical scalar and vector potentials by reconstituted Foldy-Wouthuysen transformation}

\author{Yixin Guo}
\affiliation{Department of Physics, Graduate School of Science, The University of Tokyo, Tokyo 113-0033, Japan}
\affiliation{RIKEN Nishina Center, Wako 351-0198, Japan}

\author{Haozhao Liang}
\email{haozhao.liang@riken.jp}
\affiliation{RIKEN Nishina Center, Wako 351-0198, Japan}
\affiliation{Department of Physics, Graduate School of Science, The University of Tokyo, Tokyo 113-0033, Japan}

\date{\today}

\begin{abstract}
  Inspired by the reconstituted similarity renormalization group method, the reconstituted Foldy-Wouthuysen (FW) transformation is proposed.
  Applied to the Dirac equation in the covariant density functional theory, the reconstituted FW transformation shows a fast convergence of the spectrum of the single-particle energy.
  The single-particle densities and the single-particle scalar densities obtained by this new method are also investigated.
  In particular, the relativistic corrections to the densities from the picture-change error between the Schr\"odinger and Dirac pictures are discussed in detail.
  Taking these relativistic corrections into account, both the single-particle densities and the single-particle scalar densities are almost identical to their exact values.
\end{abstract}

\maketitle

\section{Introduction}

Since the 1970s, the density functional theory (DFT) has played a crucial role in investigating the ground-state and excited-state properties of atomic nuclei in a microscopic way \cite{Serot1986, Reinhard1989Rep.Prog.Phys.52.439--514, Ring1996Prog.Part.Nucl.Phys.37.193--263, Bender2003Rev.Mod.Phys.75.121--180, Vretenar2005Phys.Rep.409.101--259, Meng2006Prog.Part.Nucl.Phys.57.470--563, Liang2015Phys.Rep.570.1--84, Meng2015J.Phys.GNucl.Part.Phys.42.093101, Nakatsukasa2016Rev.Mod.Phys.88.045004, Meng2016}.
Both the non-relativistic and the relativistic DFT have attracted a lot of attention in nuclear physics, while there still remain open questions on the connection between these two frameworks \cite{Reinhard1989Rep.Prog.Phys.52.439--514, Bender2003Rev.Mod.Phys.75.121--180, Meng2016}.

The non-relativistic expansion of the Dirac equation is considered to be a potential bridge for making the connection.
By following this direction, the Dirac equation was reduced to some non-relativistic Schr\"odinger-like equations \cite{Reinhard1989Rep.Prog.Phys.52.439--514, Bender2003Rev.Mod.Phys.75.121--180, Meng2016}.
However, on the one hand, the convergence of these methods is rather slow.
On the other hand, which is more important, the effective Hamiltonian thus obtained is not Hermitian, since the upper- or lower-component wave functions as the solutions of the Schr\"odinger-like equations alone are not orthogonal to each other.

In 2012, the similarity renormalization group (SRG) \cite{Wegner1994Ann.Phys.Berl.506.77--91, Bylev1998Phys.Lett.B428.329--333} was first applied by Guo \cite{Guo2012Phys.Rev.C85.021302} to perform the non-relativistic expansion of the single-nucleon Dirac equation for investigating the nuclear pseudospin symmetry \cite{Ginocchio1997Phys.Rev.Lett.78.436--439, Meng1999Phys.Rev.C59.154--163, Chen2003Chin.Phys.Lett.20.358--361, Zhou2003Phys.Rev.Lett.91.262501, Liang2013Phys.Rev.C87.014334,Shen2013Phys.Rev.C88.024311,Liang2015Phys.Rep.570.1--84}.
The non-relativistic expansion is carried out in the powers of ${1}/{M}$, and the results up to the ${1}/{M^3}$ order were obtained in Ref.~\cite{Guo2012Phys.Rev.C85.021302} (with $M$ the bare mass of the particle).
It is remarkable that the non-relativistic reduced Hamiltonian thus obtained is Hermitian.
In order to achieve the convergence of such a conventional SRG method, we recently showed in Ref.~\cite{Guo2019Phys.Rev.C99.054324} that the ${1}/{M^4}$-order terms are needed.
As a step further, by using the technique of resummation, which is widely used in the studies such as the Brueckner theory \cite{Brueckner1954Phys.Rev.95.217--228, Day1967Rev.Mod.Phys.39.719--744, Shen2019Prog.Part.Nucl.Phys.109.103713}, we have developed the novel reconstituted SRG method \cite{Guo2019Phys.Rev.C99.054324}, by replacing $M$ with $M^\ast$ in the non-relativistic expansion (with $M^\ast$ the Dirac mass of the particle).
Consequently, the convergence of the reconstituted SRG method became much faster than that of the conventional one.
It is worthwhile to mention that in parallel another work which started from the general operators and flow equation was carried out by Ren and Zhao \cite{Ren2019Phys.Rev.C100.044322}, where the non-relativistic expansion was also carried out in the powers of ${1}/{M^\ast}$.

In addition to the improvement of the convergence, the single-particle densities $\rho_v(\bm{r})=\psi^\dag(\bm{r})\psi(\bm{r})$, which are also called the signle-particle vector or baryon densities, calculated by the novel reconstituted SRG are also more identical to the exact values than the conventional ones \cite{Guo2019Phys.Rev.C99.054324}.
In contrast, in order to calculate the single-particle scalar densities $\rho_s(\bm{r})=\psi^\dag(\bm{r})\gamma_0\psi(\bm{r})$, the $\gamma_0$ matrix should be transformed in the same way as the Dirac Hamiltonian.
However, in both the conventional and reconstituted SRG methods, the original Dirac Hamiltonian is operated by infinite steps of unitary transformation, and it is also worthwhile to point out that a crucial superiority of the SRG technique is to avoid the treatment of the unitary transformations with specific forms.
Therefore, the calculations of the single-particle scalar densities are not trivial at all in the SRG methods.

Alternatively, the Foldy-Wouthuysen (FW) transformation, also called the Pryce-Tani-Foldy-Wouthuysen transformation \cite{Pryce1948Proc.RoyalSoc.A195.62--81, Foldy1950Phys.Rev.78.29--36,Tani1951, Foldy1952Phys.Rev.87.688--693}, which was initially formulated around 1950 to describe the relativistic spin-$1/2$ particles such as electrons in the non-relativistic limit, is also an elegant method for the non-relativistic expansion of the Dirac equation \cite{Greiner1990.277--290}.

The FW non-relativistic expansions of the Dirac equation have been carried out up to some higher orders for the selected families of potentials with specific properties, see, e.g., Refs.~\cite{Chen2010Phys.Rev.A82.012115, Chen2014Phys.Rev.A90.012112}.
Very recently, with the consideration of the strong scalar potential, we further applied the FW transformation to the general cases in the covariant DFT, and the corresponding expansion was performed up to the $1/M^4$ order \cite{Guo2019Chin.Phys.C43.114105}.
Furthermore, we also investigated the difference between the results of the conventional SRG method and the corresponding ones of the FW transformation as well as the origin of such a difference in Ref.~\cite{Guo2019Chin.Phys.C43.114105}.
As the same as the conventional SRG method, the FW transformation presents a systematic way to derive the non-relativistic expansion of the Dirac equation up to an arbitrary order.
However, when the strong scalar potential is included, the convergence of the FW transformation also becomes rather slow.
Therefore, inspired by the reconstituted SRG method, it is interesting and important to improve the FW transformation for a faster convergence.

In this paper, inspired by the reconstituted SRG method, we will propose the novel reconstituted FW transformation, where the non-relativistic expansion is performed in the powers of ${1}/{M^\ast}$.
The results up to the $1/{M^\ast}^3$ order will be shown explicitly, and the difference between these results and the corresponding ones obtained by the reconstituted SRG method will be investigated.
Furthermore, by starting generally from the field operators, the relativistic corrections to the single-particle densities will be investigated.
These corrections are originated from the so-called picture-change error between the Schr\"odinger and Dirac pictures, which was firstly applied in quantum chemistry and it was called the DKH method \cite{Douglas1974Ann.Phys.N.Y.82.89--155, Hess1986Phys.Rev.A33.3742--3748, Seino2010J.Chem.Phys.132.164108, Nakajima2012Chem.Rev.112.385--402}.
In this work, such relativistic corrections will be applied to the single-particle vector and scalar densities with the inclusion of the higher orders in a general case of the covariant DFT for the first time.

This paper is organized as follows.
The theoretical framework for the novel reconstituted FW transformation will be introduced in Sec.~\ref{sec:IIA}.
In Sec.~\ref{sec:IIB}, the single-particle density will be investigated, and as a step further, the relativistic corrections will be also considered.
The single-particle scalar density obtained by the reconstituted FW transformation and the corresponding corrections will then be discussed in Sec.~\ref{sec:IIC}.
The results for the single-particle energies and single-particle densities will be presented in Sec.~\ref{sec:IIIA} and Sec.~\ref{sec:IIIB}, respectively.
The single-particle scalar densities with the reconstituted FW transformation will be investigated in Sec.~\ref{sec:IIIC}.
Finally, a summary and perspectives will be given in Sec.~\ref{sec:IV}.

\section{Theoretical Framework}\label{sec:II}

\subsection{Reconstituted FW transformation}\label{sec:IIA}

In the relativistic scheme, the Dirac Hamiltonian with the scalar $S$ and vector $V$ potentials for nucleons reads \cite{Long2004Phys.Rev.C69.034319, Meng2006Prog.Part.Nucl.Phys.57.470--563, Zhao2010Phys.Rev.C82.054319,Liang2015Phys.Rep.570.1--84}
\begin{equation}\label{Hamiltonian}
H=\boldsymbol{\alpha} \cdot \mathbf{p}+\beta(M+S)+V,
\end{equation}
where $\alpha$ and $\beta$ are the Dirac matrices, and $M$ is the mass of nucleon.

Inspired by the reconstituted SRG method \cite{Guo2019Phys.Rev.C99.054324}, here we develop the reconstituted FW transformation.
Replacing the bare mass $M$ with the Dirac mass $M^\ast = M + S$ and considering the Hermitian property of the Hamiltonian, the corresponding operators are defined as
\begin{align}
\tilde{O}=\boldsymbol{\alpha} \cdot \mathbf{p},\quad \tilde{\varepsilon}=V,\quad \tilde{\Lambda}=-\frac{i\beta}{4}(\frac{1}{M^\ast}\tilde{O}+\tilde{O}\frac{1}{M^\ast}),
\end{align}
where the operators $\tilde{O}$ and $\tilde{\varepsilon}$ satisfy that $[\tilde{\varepsilon}, \beta] = 0$ and $\{\tilde{O}, \beta\}$ = 0, respectively.
It is important to notice that $[1/{M^\ast}, \tilde{O}] \neq 0$ here.
The Hamiltonian~\eqref{Hamiltonian} then reads
\begin{align}
H&=\beta (M+S)+\boldsymbol{\alpha} \cdot \mathbf{p}+V\nonumber\\
 &=\beta M^\ast+\tilde{O}+\tilde{\varepsilon}.
\end{align}

In order to simplify later calculations, some notations are made as follows,
\begin{subequations}
\begin{align}
\mathcal{A} &\equiv [\tilde{O},M^\ast],\\
\mathcal{B} &\equiv [\tilde{O},\frac{1}{M^\ast}],
\end{align}
and it is not difficult to derive that
\begin{align}
\frac{1}{M^\ast}\mathcal{A}+M^\ast\mathcal{B}=0.
\end{align}
\end{subequations}

Similarly to the operations in the conventional FW transforamtion \cite{Pryce1948Proc.RoyalSoc.A195.62--81, Foldy1950Phys.Rev.78.29--36, Tani1951, Foldy1952Phys.Rev.87.688--693, Greiner1990.277--290}, the Dirac Hamiltonian is transformed into
\begin{align}
H'=\,&e^{i\tilde{\Lambda}}He^{-i\tilde{\Lambda}}\nonumber\\
   =\,&H+i[\tilde{\Lambda},H]+\frac{i^2}{2!}[\tilde{\Lambda},[\tilde{\Lambda},H]]+\cdots\nonumber\\
      &+\frac{i^n}{n!}[\underbrace{\tilde{\Lambda},[\tilde{\Lambda},\cdots,[\tilde{\Lambda}}_n,H]\cdots]]+\cdots\nonumber\\
   =\,&H'_0+H'_1+\cdots+H'_n+\cdots
\end{align}
with its $n$-th component
\begin{align}
H'_n
=\frac{i^n}{n!}[\underbrace{\tilde{\Lambda},[\tilde{\Lambda},\cdots,[\tilde{\Lambda}}_n,H]\cdots]].
\end{align}

Keeping all the terms up to the $1/{M^\ast}^3$ order, the unitary transformed Hamiltonian reads
\begin{align}\label{firstH}
H'
=\,&\beta M^\ast +\tilde{\varepsilon}+\frac{1}{2}\beta \tilde{O} \frac{1}{M^\ast}\tilde{O}+\frac{1}{8}\beta[\tilde{O},\mathcal{B}] -\frac{1}{8{M^\ast}^2}[\tilde{O},[\tilde{O},\tilde{\varepsilon}]]
\nonumber\\
&-\frac{1}{8M^\ast}\mathcal{B}[\tilde{O},\tilde{\varepsilon}]
-\frac{1}{8}\beta \tilde{O}^2\frac{1}{{M^\ast}^3}\tilde{O}^2
+\frac{\beta}{2M^\ast}[\tilde{O},\tilde{\varepsilon}]\nonumber\\
&-\frac{1}{24}(4\tilde{O}^2\frac{1}{{M^\ast}^2}\tilde{O}+4\tilde{O}\frac{1}{{M^\ast}^2}\tilde{O}^2 +3\tilde{O}\frac{1}{{M^\ast}}[\tilde{O},\mathcal{B}]\nonumber\\
&+
3[\tilde{O},\mathcal{B}]\frac{1}{{M^\ast}}\tilde{O}) -\frac{1}{48{M^\ast}^3}\beta[\tilde{O},[\tilde{O},[\tilde{O},\tilde{\varepsilon}]]].
\end{align}

In this unitary transformed Hamiltonian~(\ref{firstH}), the off-diagonal parts are not zero but raised by one order from $\tilde{O}$ to $\frac{1}{2M^\ast}\beta[\tilde{O},\tilde{\varepsilon}]$.
In order to make the off-diagonal parts higher than a given order (e.g., the $1/{M^\ast}^3$ order), one should repeat the FW transformation until the accuracy is achieved \cite{Greiner1990.277--290, Guo2019Chin.Phys.C43.114105}.
For that, the operators $\tilde{O}'$ and $\tilde{\varepsilon}'$ are redefined according to Eq.~(\ref{firstH}), and one has
\begin{subequations}
\begin{align}
\tilde{\varepsilon}' = \,&\tilde{\varepsilon}+\frac{1}{2}\beta \tilde{O} \frac{1}{M^\ast}\tilde{O}-\frac{1}{8{M^\ast}^2}[\tilde{O},[\tilde{O},\tilde{\varepsilon}]] +\frac{1}{8}\beta[\tilde{O},\mathcal{B}]\nonumber\\
&-\beta \tilde{O}^2\frac{1}{8{M^\ast}^3}\tilde{O}^2
-\frac{1}{8M^\ast}\mathcal{B}[\tilde{O},\tilde{\varepsilon}],\\
\tilde{O}' = \,&\frac{\beta}{2M^\ast}[\tilde{O},\tilde{\varepsilon}]-\frac{1}{24}\Big(4\tilde{O}^2\frac{1}{{M^\ast}^2}\tilde{O} +4\tilde{O}\frac{1}{{M^\ast}^2}\tilde{O}^2
          \nonumber\\
&+3\tilde{O}\frac{1}{{M^\ast}}[\tilde{O},\mathcal{B}]+
3[\tilde{O},\mathcal{B}]\frac{1}{{M^\ast}}\tilde{O}\Big)\nonumber\\
&-\frac{1}{48{M^\ast}^3}\beta[\tilde{O},[\tilde{O},[\tilde{O},\tilde{\varepsilon}]]],\\
    \tilde{\Lambda}' = \,&-\frac{i\beta}{4}\left(\frac{1}{M^\ast}\tilde{O}'+\tilde{O}'\frac{1}{M^\ast}\right).
\end{align}
\end{subequations}
The corresponding reconstituted FW transformation reads
\begin{align}
H''= e^{i\tilde{\Lambda}'} H' e^{-i\tilde{\Lambda}'}.
\end{align}

With the recursions \cite{Guo2019Chin.Phys.C43.114105}, the Hamiltonian of the reconstituted FW transformation eventually reads
\begin{align}\label{FWH}
\mathcal{H}_{\rm rFW}=\,&
\beta M^\ast +\tilde{\varepsilon}
+\frac{1}{2}\beta \tilde{O} \frac{1}{M^\ast}\tilde{O}
+\frac{1}{8}\beta[\tilde{O},\mathcal{B}]\nonumber\\
&-\frac{1}{8{M^\ast}^2}[\tilde{O},[\tilde{O},\tilde{\varepsilon}]]
-\frac{1}{8M^\ast}\mathcal{B}[\tilde{O},\tilde{\varepsilon}]\nonumber\\
&-\frac{1}{8}\beta \tilde{O}^2\frac{1}{{M^\ast}^3}\tilde{O}^2
-\frac{1}{8{M^\ast}^3}\beta[\tilde{O},\tilde{\varepsilon}][\tilde{O},\tilde{\varepsilon}],
\end{align}
and its off-diagonal parts have been raised up to at least the $1/{M^\ast}^4$ order.

For the systems with the spherical symmetry, the corresponding radial single-nucleon Dirac equation reads \cite{Meng2006Prog.Part.Nucl.Phys.57.470--563, Liang2015Phys.Rep.570.1--84}
\begin{equation}\label{eq:Dirac1}
\left(
\begin{array}{cc}
\Sigma(r)+M & -\frac{\textrm{d}}{\textrm{d}r}+\frac{\kappa}{r}\\
\frac{\textrm{d}}{\textrm{d}r}+\frac{\kappa}{r} & \Delta(r)-M
\end{array}
\right )
\left(
\begin{array}{c}
G(r) \\ F(r)
\end{array}
\right )
=E
\left(
\begin{array}{c}
G(r) \\ F(r)
\end{array}
\right ),
\end{equation}
with the normalization condition $\int \mathrm{d}r\,[G^2(r) + F^2(r)] = 1$,
where $\kappa$ is a good quantum number defined as $\kappa=\mp~(j+{1}/{2})$ for $j=l\pm{1}/{2}$, and $\Sigma(r) = V(r) + S(r)$ and $\Delta(r) = V(r) - S(r)$ are the sum of and the difference between the vector and scalar potentials, respectively.
The single-particle energies $E = \varepsilon +M $ include the mass of nucleon.
The initial conditions read
\begin{align}\label{initialnew}
\tilde{\varepsilon}={\left( \begin{array}{cc}
V(r) & 0\\
0 & V(r)
\end{array}
\right )}&,\quad
\tilde{O}={\left( \begin{array}{cc}
0 &-\frac{\textrm{d}}{\textrm{d}r}+\frac{\kappa}{r}\\
\frac{\textrm{d}}{\textrm{d}r}+\frac{\kappa}{r}& 0
\end{array}
\right )},\nonumber\\
\mathcal{A}={\left( \begin{array}{cc}
0 & -{M^\ast}'\\
{M^\ast}' & 0
\end{array}
\right )}&,\quad
\mathcal{B}={\left( \begin{array}{cc}
0 & \frac{{M^\ast}'}{{M^\ast}^2}\\
-\frac{{M^\ast}'}{{M^\ast}^2} & 0
\end{array}
\right )}.
\end{align}

According to Eq.~(\ref{FWH}), the Dirac Hamiltonian is transformed by the reconstituted FW transformation as
\begin{equation}\label{eq:HFW}
{\left( \begin{array}{cc}
\mathcal{H}^{\rm (F)}_{\rm rFW} + M & O(\frac{1}{{M^\ast}^4}) \\
O(\frac{1}{{M^\ast}^4}) & \mathcal{H}^{\rm (D)}_{\rm rFW} - M
\end{array}
\right)}.
\end{equation}

Hereafter, we will focus on the single-particle states in the Fermi sea, which correspond to their counterparts in the non-relativistic framework.
Therefore, $\mathcal{H}^{\rm (F)}_{\rm rFW}$ will be investigated in detail, and its superscript will be omitted when there is no confusion.

The expansion of $\mathcal{H}_{\rm rFW}$ up to the $1/{M^\ast}^3$ order is carefully worked out as
\begin{subequations}\label{neq}
\begin{align}
\mathcal{H}_{\rm rFW,0}=\,&\Sigma,\label{neq1}\\
\mathcal{H}_{\rm rFW,1}=\,&-\frac{\textrm{d}}{\textrm{d}r}\frac{1}{2M^\ast}\frac{\textrm{d}}{\textrm{d}r}+\frac{\kappa(\kappa+1)}{2M^\ast r^2},\label{neq2}\\
\mathcal{H}_{\rm rFW,2}=\,&-\frac{\Delta'}{4{M^\ast}^2}\frac{\kappa}{r}+\frac{\Sigma''}{8{M^\ast}^2},\label{neq3}\\
\mathcal{H}_{\rm rFW,3}=\,&-p^2\frac{1}{8{M^\ast}^3}p^2 +\frac{\Sigma'^2}{8{M^\ast}^3}-\frac{3S'\Sigma'}{8{M^\ast}^3},\label{neq4}
\end{align}
\end{subequations}
where
\begin{equation}
p^2=-\frac{\textrm{d}^2}{\textrm{d}r^2}+\frac{\kappa(\kappa+1)}{r^2}.
\end{equation}

It is noted that operators with higher-order derivatives appear from $\mathcal{H}_{\rm rFW,3}$, and thus the eigenequation containing up to the second derivatives reads
\begin{equation}\label{eq:Sch2}
  \left[\mathcal{H}_{\rm rFW,0} + \mathcal{H}_{\rm rFW,1} + \mathcal{H}_{\rm rFW,2}\right] \varphi_k(r) = \varepsilon_k\varphi_k(r),
\end{equation}
with the normalization condition $\int \mathrm{d}r\,\varphi^2(r) = 1$.
Following the same way in Ref.~\cite{Guo2019Phys.Rev.C99.054324}, the eigenequation~(\ref{eq:Sch2}) will be solved and the higher-order term $\mathcal{H}_{\rm rFW,3}$ will be calculated by the perturbation theory in the following discussions.

In Ref.~\cite{Guo2019Chin.Phys.C43.114105}, the relation between the conventional SRG method and FW transformation has been investigated.
Similarly, by comparing Eqs.~(\ref{neq1}), (\ref{neq2}), (\ref{neq3}) and (\ref{neq4}) with the corresponding results obtained in the non-relativistic expansion by the reconstituted SRG method \cite{Guo2019Phys.Rev.C99.054324}, one finds the differences as
\begin{subequations}\label{eq:diff}
\begin{align}
\mathcal{H}_{\rm rFW,0}-\mathcal{H}_{\rm rSRG,0}=\,&0\\
\mathcal{H}_{\rm rFW,1}-\mathcal{H}_{\rm rSRG,1}=\,&0\\
\mathcal{H}_{\rm rFW,2}-\mathcal{H}_{\rm rSRG,2}=\,&0\\
\mathcal{H}_{\rm rFW,3}-\mathcal{H}_{\rm rSRG,3}=\,&\frac{\Delta'\Sigma'}{16{M^\ast}^3}.\label{neq5}
\end{align}
\end{subequations}

It turns out that the differences shown in Eq.~\eqref{eq:diff} come from an additional unitary transformation, after the Dirac Hamiltonian is decoupled into the upper and lower parts.
Let
\begin{align}
\Xi=\,&-\frac{i\beta}{64}\left(\frac{1}{{M^\ast}^3}[\tilde{O}^2,\tilde{\varepsilon}-\beta{M^\ast}]+[\tilde{O}^2,\tilde{\varepsilon}-\beta{M^\ast}]
\frac{1}{{M^\ast}^3}\right)\nonumber\\
&+\cdots
\end{align}
It is a Hermitian and diagonal operator, i.e., $\Xi^\dag = \Xi$ and $\beta\Xi = \Xi\beta$.
Acting an additional unitary transformation on $\mathcal{H}_{\rm rFW}$, it reads
\begin{align}
e^{i\Xi}\mathcal{H}_{\rm rFW}e^{-i\Xi}=\mathcal{H}_{\rm rFW}+i[\Xi,\mathcal{H}_{\rm rFW}]+\cdots
\end{align}
\begin{widetext}
Keeping all the terms up to the $1/{M^\ast}^3$ order, the results are
\begin{align}\label{neq6}
\mathcal{H}_{\rm rFW}+\frac{\beta}{64}\Big[\frac{1}{{M^\ast}^3}[\tilde{O}^2,\tilde{\varepsilon}-\beta{M^\ast}]
+[\tilde{O}^2,\tilde{\varepsilon}-\beta{M^\ast}]\frac{1}{{M^\ast}^3},\beta M^\ast+\tilde{\varepsilon}\Big].
\end{align}
As a step further, it is not complicated to obtain the upper-left component of Eq.~\eqref{neq6} as
\begin{align}
\mathcal{H}_{\rm rFW,0}+\mathcal{H}_{\rm rFW,1}+\mathcal{H}_{\rm rFW,2}+\mathcal{H}_{\rm rFW,3}-\frac{\Delta'\Sigma'}{16{M^\ast}^3},
\end{align}
which is nothing but $\mathcal{H}_{\rm rSRG}$.
\end{widetext}

Since $e^{-i\Xi}$ is a unitary operator acting on the already decoupled Hamiltonian, it does not affect the non-relativistic expansion of the Dirac equation, in the sense that the single-particle spectrum obtained by the reconstituted FW transformation is identical to that obtained by the reconstituted SRG method.
At the same time, the above conclusions provide the foundation of the reconstituted SRG method proposed in Ref.~\cite{Guo2019Phys.Rev.C99.054324}.

\subsection{Single-particle density}\label{sec:IIB}

The relativistic single-particle wave-function $|\psi\rangle$ satisfies
\begin{align}
H|\psi\rangle=E|\psi\rangle,
\end{align}
and the non-relativistic counterpart $|\varphi\rangle$ in the FW transformation satisfies
\begin{align}
\mathcal{H}_{\rm FW}|\varphi\rangle=E|\varphi\rangle
\end{align}
with
\begin{align}
\mathcal{H}_{\rm FW}=e^{i{\Lambda_{\rm FW}}}He^{-i{\Lambda_{\rm FW}}}
\end{align}
and
\begin{align}
|\varphi\rangle=e^{i{\Lambda_{\rm FW}}}|\psi\rangle.
\end{align}

Since only the unitary transformations have been applied in the FW transformation, it seems naturally that the single-particle density $\rho_v(\bm{r})=\psi^\dag(\bm{r})\psi(\bm{r})$ will be calculated as
\begin{align}\label{rhov1}
\rho_{v, \textrm{non}}(r)
=\frac{1}{4\pi r^2}\varphi^\dag(r)\varphi(r).
\end{align}
However, by checking the details, it can be found that
\begin{align}
\rho_{v,\textrm{exact}}(r)=\rho_{v,\textrm{non}}(r)+\Delta\rho_v(r),
\end{align}
where $\Delta\rho_v$ actually comes from the so-called picture-change error \cite{Douglas1974Ann.Phys.N.Y.82.89--155, Hess1986Phys.Rev.A33.3742--3748}, i.e., the difference between the densities calculated in the Schr\"odinger picture and the ones obtained in the Dirac picture.

In order to investigate the correction to the single-particle density explicitly, we start from its strict definition, which reads
\begin{align}
\rho_v(\bm{r})=\,&\langle\psi|\hat{c}^\dag(\bm{r})\hat{c}(\bm{r})|\psi\rangle\nonumber\\
=\,&\int\, \textrm{d}\bm{r}'\psi^\dag(\bm{r}')\delta(\bm{r}'-\bm{r})\psi(\bm{r}').
\end{align}
Here $\hat{c}^\dag(\bm{r})$ and $\hat{c}(\bm{r})$ are the field operators (i.e., the particle creation and annihilation operators with the index of spatial coordinate).
Therefore, the single-particle density reads
\begin{align}
\rho_v(\bm{r})=\,&\langle\varphi|e^{i{\Lambda_{\rm FW}}}\hat{c}^\dag(\bm{r})\hat{c}(\bm{r})e^{-i{\Lambda_{\rm FW}}}|\varphi\rangle\nonumber\\
=\,&\int\,\textrm{d}\bm{r}'\varphi^\dag(\bm{r}')e^{i{\Lambda_{\rm FW}}(\bm{r}')}\delta(\bm{r}'-\bm{r})e^{-i{\Lambda_{\rm FW}}(\bm{r}')}\varphi(\bm{r}').
\end{align}
Hereafter, the delta function $\delta(\bm{r}'-\bm{r})$ will be denoted as $\boldsymbol{\Delta}(\bm{r}',\bm{r})$  (with a bold font in order to distinguish from $\Delta=V-S$).

In the present scheme, all operators should be transformed in the same way as the Dirac Hamiltonian.
Therefore, the delta function in the FW formalism is transformed as
\begin{align}
\boldsymbol{\Delta}'=e^{i\tilde{\Lambda}}\boldsymbol{\Delta}e^{-i\tilde{\Lambda}}.
\end{align}
The expansion of $\boldsymbol{\Delta}'$ reads
\begin{align}
\boldsymbol{\Delta}'
  =\,&\boldsymbol{\Delta}+i[\tilde{\Lambda},\boldsymbol{\Delta}]+\frac{i^2}{2!}[\tilde{\Lambda},[\tilde{\Lambda},\boldsymbol{\Delta}]]\nonumber\\
   &+\cdots+\frac{i^n}{n!}[\underbrace{\tilde{\Lambda},
   [\tilde{\Lambda},\cdots,[\tilde{\Lambda}}_n,\boldsymbol{\Delta}]\cdots]]+\cdots
\end{align}
with its $n$-th component
\begin{align}
\boldsymbol{\Delta}'_n
=\frac{i^n}{n!}[\underbrace{\tilde{\Lambda},[\tilde{\Lambda},\cdots,[\tilde{\Lambda}}_n,\boldsymbol{\Delta}]\cdots]].
\end{align}
As a result,
\begin{align}
\boldsymbol{\Delta}'=\,&\boldsymbol{\Delta}-\frac{1}{8{M^\ast}^2}\boldsymbol{\Delta}^{(2)}-\frac{1}{8M^\ast}\mathcal{B}\boldsymbol{\Delta}^{(1)}+\frac{\beta}{2M^\ast}\boldsymbol{\Delta}^{(1)}\nonumber\\
&-\frac{\beta}{48{M^\ast}^3}\boldsymbol{\Delta}^{(3)}.
\end{align}
Here the relevant notations are defined as
\begin{subequations}
\begin{align}
\boldsymbol{\Delta}^{(1)} &\equiv [\tilde{O},\boldsymbol{\Delta}],\\
\boldsymbol{\Delta}^{(2)} &\equiv [\tilde{O},\boldsymbol{\Delta}^{(1)}],\\
\boldsymbol{\Delta}^{(3)} &\equiv [\tilde{O},\boldsymbol{\Delta}^{(2)}].
\end{align}
\end{subequations}

Following the recursion relation in the reconstituted FW transformation,
the result up to the $1/{M^\ast}^3$ order reads
\begin{align}
\boldsymbol{\Delta}_{\rm rFW}
=\,&\boldsymbol{\Delta}-\frac{1}{8{M^\ast}^2}\boldsymbol{\Delta}^{(2)}-\frac{1}{8M^\ast}\mathcal{B}\boldsymbol{\Delta}^{(1)}-\frac{\beta}{4{M^\ast}^3}[\tilde{O},\tilde{\varepsilon}]\boldsymbol{\Delta}^{(1)}\nonumber\\
\,&+\frac{\beta}{2M^\ast}\boldsymbol{\Delta}^{(1)}-\frac{\beta}{2{M^\ast}^3}\boldsymbol{\Delta}^{(1)}\tilde{O}^2-\frac{\beta}{2{M^\ast}^3}\boldsymbol{\Delta}^{(2)}\tilde{O}\nonumber\\
&-\frac{9\beta}{48{M^\ast}^3}\boldsymbol{\Delta}^{(3)}.
\end{align}

The delta function can be expanded in terms of spherical harmonics,
\begin{align}
\delta(\bm{r}'-\bm{r})=\frac{1}{r^2}\delta(r'-r)\sum_{L=0}^{\infty}\bm{Y}_L(\hat{\bm{r}}')\cdot\bm{Y}_L(\hat{\bm{r}}).
\end{align}
When the spherical symmetry is taken into account, it reads
\begin{align}
\delta(\bm{r}'-\bm{r})=\frac{1}{4\pi r^2}\delta(r'-r).
\end{align}
Thus,
\begin{align}
\boldsymbol{\Delta}(r',r)=
\frac{1}{4\pi r^2}
\left( \begin{array}{cc}
\delta(r'-r) & 0\\
0 & \delta(r'-r)
\end{array}
\right ).
\end{align}
\begin{widetext}
In addition, one has
\begin{subequations}
\begin{align}
\boldsymbol{\Delta}^{(1)}(r',r)=\,&
\frac{1}{4\pi r^2}
\left( \begin{array}{cc}
0 & -\delta'\\
\delta' & 0
\end{array}
\right ),\\
\boldsymbol{\Delta}^{(2)}(r',r)=\,&
\frac{1}{4\pi r^2}
\left( \begin{array}{cc}
-\delta''+\frac{2\kappa}{r'}\delta' & 0\\
0 & -\delta''-\frac{2\kappa}{r'}\delta'
\end{array}
\right ),\\
\boldsymbol{\Delta}^{(3)}(r',r)=\,&
\frac{1}{4\pi r^2}
\left( \begin{array}{cc}
0& \delta'''-\frac{2\kappa}{r'}\delta''+ \frac{2\kappa}{r'^2}\delta'\\
-\delta'''-\frac{2\kappa}{r'}\delta''+ \frac{2\kappa}{r'^2}\delta' & 0
\end{array}
\right ).
\end{align}
\end{subequations}
Here the short-hand writings mean
\begin{align}
\delta' =\frac{\textrm{d}}{\textrm{d}r'}\delta(r'-r),\qquad
\delta'' =\frac{\textrm{d}^2}{\textrm{d}r'^2}\delta(r'-r),\qquad
\delta''' =\frac{\textrm{d}^3}{\textrm{d}r'^3}\delta(r'-r).
\end{align}

The meanings of the derivatives of delta function are related to the integrals by part.
In detail, for arbitrary wave functions $\eta(r')$ and $\xi(r')$, the integral
\begin{align}
I(r)= \int\,\eta^\dag(r')\frac{\textrm{d}\delta(r'-r)}{\textrm{d}r'}\xi(r')\textrm{d}r'
= \left.-\eta^\dag(r') \left(\overleftarrow{\frac{\textrm{d}}{\textrm{d}r'}} +\overrightarrow{\frac{\textrm{d}}{\textrm{d}r'}}\right)\xi(r')\right|_{r'=r}
= -\eta^\dag(r) \left(\overleftarrow{\frac{\textrm{d}}{\textrm{d}r}} +\overrightarrow{\frac{\textrm{d}}{\textrm{d}r}}\right)\xi(r),
\end{align}
where the arrow $\leftarrow$ ($\rightarrow$) means the operator acts only on the functions on its left-hand (right-hand) side.
It is important to point out that the action of taking $r'=r$ should be performed after the derivatives in order to avoid the mistakes which may be caused by the   confusion of $r$ and $r'$ during the calculations.
This indicates that the first derivative of delta function can be formally expressed as
\begin{align}
\int\,\textrm{d}r'\cdots\frac{\textrm{d}\delta(r'-r)}{\textrm{d}r'}\cdots
\quad = \quad -\cdots\left(\overleftarrow{\frac{\textrm{d}}{\textrm{d}r}}+\overrightarrow{\frac{\textrm{d}}{\textrm{d}r}}\right)\cdots
\end{align}
Similarly, the second and third derivatives of delta function correspond to
\begin{align}
\int\,\textrm{d}r'\cdots\frac{\textrm{d}^2\delta(r'-r)}{\textrm{d}r'^2}\cdots
\quad = \quad
 \cdots\left(\overleftarrow{\frac{\textrm{d}^2}{\textrm{d}r^2}}+2\overleftarrow{\frac{\textrm{d}}{\textrm{d}r}}
\overrightarrow{\frac{\textrm{d}}{\textrm{d}r}}+\overrightarrow{\frac{\textrm{d}^2}{\textrm{d}r^2}}\right)\cdots
\end{align}
and
\begin{align}
\int\,\textrm{d}r'\cdots\frac{\textrm{d}^3\delta(r'-r)}{\textrm{d}r'^3}\cdots
\quad = \quad
-\cdots\left(\overleftarrow{\frac{\textrm{d}^3}{\textrm{d}r^3}}+3\overleftarrow{\frac{\textrm{d}^2}{\textrm{d}r^2}}
\overrightarrow{\frac{\textrm{d}}{\textrm{d}r}}+3\overleftarrow{\frac{\textrm{d}}{\textrm{d}r}}\overrightarrow{\frac{\textrm{d}^2}{\textrm{d}r^2}}
+\overrightarrow{\frac{\textrm{d}^3}{\textrm{d}r^3}}\right)\cdots
\end{align}
respectively.

Finally, the delta function with the reconstituted FW transformation reads
\begin{align}\label{eq:deltaFW}
4\pi r^2\boldsymbol{\Delta}_{\rm rFW}(r,r)=\,&
1 + \frac{1}{8{M^\ast}^2}\left(\overleftarrow{\frac{\textrm{d}^2}{\textrm{d}r^2}} +2\overleftarrow{\frac{\textrm{d}}{\textrm{d}r}}\overrightarrow{\frac{\textrm{d}}{\textrm{d}r}} +\overrightarrow{\frac{\textrm{d}^2}{\textrm{d}r^2}}\right)
+\frac{\beta}{4{M^\ast}^2}\frac{\kappa}{r}\left(\overleftarrow{\frac{\textrm{d}}{\textrm{d}r}} +\overrightarrow{\frac{\textrm{d}}{\textrm{d}r}}\right)\nonumber\\
&+\frac{S'}{8{M^\ast}^3}\left(\overleftarrow{\frac{\textrm{d}}{\textrm{d}r}} +\overrightarrow{\frac{\textrm{d}}{\textrm{d}r}}\right)
-\frac{\beta V'}{4{M^\ast}^3}\left(\overleftarrow{\frac{\textrm{d}}{\textrm{d}r}} +\overrightarrow{\frac{\textrm{d}}{\textrm{d}r}}\right)\nonumber\\
&+\frac{\gamma_5}{2M^\ast}\left(\overleftarrow{\frac{\textrm{d}}{\textrm{d}r}} +\overrightarrow{\frac{\textrm{d}}{\textrm{d}r}}\right)
+\left( \begin{array}{cc}
0& O(\frac{1}{{M^\ast}^3})\\
O(\frac{1}{{M^\ast}^3}) & 0
\end{array}
\right),
\end{align}
where $\gamma_5$ is the Dirac matrix.

Consequently, after considering these corrections from the FW transformation, which we call the relativistic corrections, the single-particle density for the systems with the spherical symmetry reads
\begin{align}\label{rhov2}
4\pi r^2\rho_v(r)
=\,&
\left.\left( \begin{array}{cc}
\varphi^\dag(r')& 0
\end{array}
\right )
\boldsymbol{\Delta}_{\rm rFW}(r',r)
\left( \begin{array}{c}
\varphi(r')\\ 0
\end{array}
\right )\right|_{r'=r}\nonumber\\
=\,&\varphi^\dag(r)\left[1-\left(-\overleftarrow{\frac{\textrm{d}^2}{\textrm{d}r^2}}
+\frac{\kappa(\kappa+1)}{r^2}\right)\frac{1}{8{M^\ast}^2}
-\frac{1}{8{M^\ast}^2}\left(-\overrightarrow{\frac{\textrm{d}^2}{\textrm{d}r^2}} +\frac{\kappa(\kappa+1)}{r^2}\right)
+\left(\overleftarrow{\frac{\textrm{d}}{\textrm{d}r}}
+\frac{\kappa}{r}\right)\frac{1}{4{M^\ast}^2}\left(\overrightarrow{\frac{\textrm{d}}
{\textrm{d}r}}+\frac{\kappa}{r}\right)\right.\nonumber\\
&\qquad\qquad\left.+\overleftarrow{\frac{\textrm{d}}{\textrm{d}r}}\left(\frac{-3S'-2V'}{8{M^\ast}^3}\right) +\left(-\frac{S'}{2{M^\ast}^3}\frac{\kappa}{r}
-\frac{S''+2V''}{8{M^\ast}^3}\right)
+\left(\frac{-3S'-2V'}{8{M^\ast}^3}\right)\overrightarrow{\frac{\textrm{d}}{\textrm{d}r}}\right]\varphi(r).
\end{align}
\end{widetext}

\subsection{Single-particle scalar density}\label{sec:IIC}

For the single-particle scalar density $\rho_s(\bm{r})=\psi^\dag(\bm{r})\gamma_0\psi(\bm{r})$, we need to consider the FW transformations of not only the delta function as shown in Eq.~\eqref{eq:deltaFW} but also the Dirac matrix $\gamma_0$.
By following the procedures in the previous sections, the Dirac matrix $\gamma_0$ with the reconstituted FW transformation reads
\begin{align}
\tilde{\gamma}=\,&
\gamma_0-\frac{1}{4M^\ast} \beta[\tilde{O},\mathcal{B}] -\frac{1}{2}\beta\tilde{O}\frac{1}{{M^\ast}^2}\tilde{O} +\frac{1}{4{M^\ast}^3}[\tilde{O},[\tilde{O},\tilde{\varepsilon}]]\nonumber\\
&-\frac{1}{2}\left(\frac{1}{M^\ast}\tilde{O} +\tilde{O}\frac{1}{M^\ast}\right) -\frac{1}{2{M^\ast}^2}\beta[\tilde{O},\tilde{\varepsilon}]\nonumber\\
&+\frac{1}{4}\left(\tilde{O}^2\frac{1}{{M^\ast}^3}\tilde{O} +\tilde{O}\frac{1}{{M^\ast}^3}\tilde{O}^2\right) -\frac{1}{4{M^\ast}^3}[[\tilde{O},\tilde{\varepsilon}],\tilde{\varepsilon}],
\end{align}
up to the $1/{M^\ast}^3$ order.
It can be expressed as a two-by-two matrix
\begin{align}
\tilde{\gamma}=\left( \begin{array}{cc}
\tilde{\gamma}_{11} & \tilde{\gamma}_{12}\\
\tilde{\gamma}_{21} & \tilde{\gamma}_{22}
\end{array}
\right ),
\end{align}
and with the spherical symmetry
its matrix elements read
\begin{subequations}
\begin{align}
\tilde{\gamma}_{11}=\,&1+\frac{\textrm{d}}{\textrm{d}r}\frac{1}{2{M^\ast}^2}\frac{\textrm{d}}{\textrm{d}r}-\frac{\kappa(\kappa+1)}{2{M^\ast}^2 r^2}
+\frac{\Delta'}{2{M^\ast}^3}\frac{\kappa}{r}\nonumber\\
&-\frac{\Sigma''}{4{M^\ast}^3}+O(\frac{1}{{M^\ast}^4}),\label{gamma11}\\
\tilde{\gamma}_{12}=\,&\frac{1}{2}\left(\frac{1}{M^\ast}\frac{\textrm{d}}{\textrm{d}r}+\frac{\textrm{d}}{\textrm{d}r}\frac{1}{M^\ast}\right)-\frac{1}{M^\ast}\frac{\kappa}{r}+\frac{V'}{2{M^\ast}^2}\nonumber\\
&+O(\frac{1}{{M^\ast}^3}),\\
\tilde{\gamma}_{21}=\,&-\frac{1}{2}\left(\frac{1}{M^\ast}\frac{\textrm{d}}{\textrm{d}r}+\frac{\textrm{d}}{\textrm{d}r}\frac{1}{M^\ast}\right)-\frac{1}{M^\ast}\frac{\kappa}{r}+\frac{V'}{2{M^\ast}^2}\nonumber\\
&+O(\frac{1}{{M^\ast}^3}).
\end{align}
\end{subequations}

\begin{widetext}
Up to here, with the reconstituted FW transformation, the scalar densities in the Schr\"odinger picture are in the forms of
\begin{align}\label{rhos1}
4\pi r^2\rho_{s,\textrm{non}}(r)
=\varphi^\dag(r)
\left(1+\frac{\textrm{d}}{\textrm{d}r}\frac{1}{2{M^\ast}^2}\frac{\textrm{d}}{\textrm{d}r}-\frac{\kappa(\kappa+1)}{2{M^\ast}^2 r^2}
+\frac{\Delta'}{2{M^\ast}^3}\frac{\kappa}{r}-\frac{\Sigma''}{4{M^\ast}^3}\right)\varphi(r).
\end{align}

As a step further, similarly to the single-particle density, the single-particle scalar density can be modified with the consideration of the relativistic corrections originated from the picture-change error between the Schr\"odinger and Dirac pictures, i.e.,
\begin{align}
\rho_{s,\textrm{exact}}(r)=\rho_{s,\textrm{non}}(r)+ \Delta \rho_s(r).
\end{align}
It is important to point out that the difference between $\rho_{v,\textrm{non}}$ and $\rho_{s,\textrm{non}}$ appears from the $1/{M^\ast}^2$ order, and the relativistic correction $\Delta \rho_s$ also appears from the same order.
Therefore, the relativistic corrections will also play a crucial role in the difference between the vector and scalar densities.

For the systems with the spherical symmetry, the single-particle scalar density reads
\begin{align}\label{rhos2}
4\pi r^2\rho_s(r)
=\,&
\left.\left( \begin{array}{cc}
\varphi^\dag(r')& 0
\end{array}
\right)
\boldsymbol{\Delta}_{\rm rFW}(r',r)\,\tilde{\gamma}\,
\left( \begin{array}{c}
\varphi(r')\\ 0
\end{array}
\right )\right|_{r'=r}\nonumber\\
=\,&
\varphi^\dag(r)\left[
1-\left(-\overleftarrow{\frac{\textrm{d}^2}{\textrm{d}r^2}}+\frac{\kappa(\kappa+1)}{r^2}\right)\frac{1}{8{M^\ast}^2}
-\frac{1}{8{M^\ast}^2}\left(-\overrightarrow{\frac{\textrm{d}^2}{\textrm{d}r^2}} +\frac{\kappa(\kappa+1)}{r^2}\right)
-\left(\overleftarrow{\frac{\textrm{d}}{\textrm{d}r}}+\frac{\kappa}{r}\right)\frac{1}{4{M^\ast}^2}
\left(\overrightarrow{\frac{\textrm{d}}{\textrm{d}r}}+\frac{\kappa}{r}\right)\right.\nonumber\\
&\qquad\qquad\left.-\overleftarrow{\frac{\textrm{d}}{\textrm{d}r}}\frac{S'}{8{M^\ast}^3} +\frac{V'}{2{M^\ast}^3}\frac{\kappa}{r} -\frac{S''+2V''}{8{M^\ast}^3}
-\frac{S'}{8{M^\ast}^3}\overrightarrow{\frac{\textrm{d}}{\textrm{d}r}}\right]\varphi(r).
\end{align}
In this expression, there are also contributions from the off-diagonal parts of $\boldsymbol{\Delta}_{\rm rFW}(r,r)$ and $\tilde{\gamma}$.
\end{widetext}

\section{Results and Discussion}\label{sec:III}

In order to make the comparisons and discussions with the corresponding results obtained by the SRG methods in Ref.~\cite{Guo2019Phys.Rev.C99.054324}, we use the Woods-Saxon forms for the scalar and vector potentials, $\Sigma(r)=\Sigma_0f(a_0,r_0,r)$ and $\Delta(r)=\Delta_0f(a_0,r_0,r)$, with
\begin{equation}
f(a_0,r_0,r)=\frac{1}{1+e^{\frac{r-r_0}{a_0}}},
\end{equation}
where the parameters are the same as those in Ref.~\cite{Guo2019Phys.Rev.C99.054324}, i.e., $\Sigma_0=-66.0$~MeV, $\Delta_0 = 650.0$~MeV, $r_0= 7.0$~fm, and $a_0= 0.6$~fm, which are the typical values for neutrons in the nucleus $^{208}\textrm{Pb}$.
The mass of nucleon is taken as $M=939.0$~MeV.

The Dirac equation~(\ref{eq:Dirac1}) is solved in coordinate space by the shooting method \cite{Meng1998Nucl.Phys.A635.3--42} within a spherical box with radius $R_{\rm box}= 20$~fm and mesh size $\textrm{d}r = 0.05$~fm.
The single-particle energies and densities thus obtained will serve as benchmarks, labelled as Exact in the tables and figures.
The non-relativistic equation~(\ref{eq:Sch2}) is also solved in coordinate space by the shooting method with the same box and mesh sizes.

\subsection{Single-particle spectrum}\label{sec:IIIA}

\begin{figure}
  \includegraphics[width=0.45\textwidth]{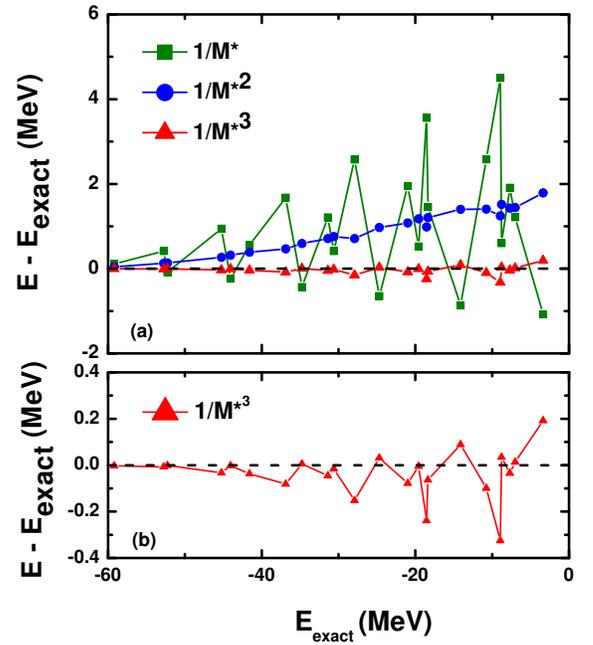}
  \caption{(Color online) (a) Discrepancy between the neutron single-particle energies obtained by the reconstituted FW transformation and the exact ones as a function of the single-particle energy.
  The squares, circles, and triangles represent the energy discrepancy at the $1/M^\ast$, $1/{M^\ast}^2$, and $1/{M^\ast}^3$ orders for the reconstituted FW transformation, respectively.
  (b) Discrepancy at the $1/{M^\ast}^3$ order on a smaller scale.}\label{fig1}
\end{figure}

In Fig.~\ref{fig1}, the discrepancy between the neutron single-particle energies obtained by the reconstituted FW transformation and the exact ones is shown as a function of the single-particle energy.
In the present case of $^{208}$Pb, the last occupied orbital is the $3p_{1/2}$ state, whose energy is $-6.999$~MeV.
It can be seen that the non-relativistic expansion with the reconstituted FW transformation achieves a fast convergence.
Shown in a smaller scale in Panel (b) of Fig.~\ref{fig1}, the differences between the single-particle energies obtained at the $1/{M^\ast}^3$ order and the exact ones are less than $0.1$~MeV for the deeply bound states, and become slightly larger to around $0.2$~MeV for the weakly bound states.
The three orbitals that show the biggest negative differences are the $1g_{9/2}$ state at $-28.072$~MeV, $1h_{11/2}$ state at $-18.784$~MeV, and $1i_{13/2}$ state at $-9.267$~MeV.
Their deviations from the corresponding exact values are $-152$, $-239$, and $-325$~keV, respectively.
In contrast, the orbital that shows the biggest positive difference is the $1h_{9/2}$ state at $-14.027$~MeV, whose deviation from the corresponding exact value is $89$~keV.

In order to understand these energy differences for the states with large angular momenta $l$, we recall that, although a discrepancy between the results of the conventional SRG method \cite{Guo2019Phys.Rev.C99.054324} and FW transformation \cite{Guo2019Chin.Phys.C43.114105} appears starting from the $1/M^3$ order, their spectrum of the single-particle energies are identical to each other \cite{Guo2019Chin.Phys.C43.114105}.
Thus the detailed $1/M^4$-order results of the conventional SRG method derived in Ref.~\cite{Guo2019Phys.Rev.C99.054324} can help us investigate the results in Fig.~\ref{fig1} more deeply here.
It can be seen that except the terms that have already included so far with the replacement of $M$ by $M^\ast$, the three biggest contributions in the $1/M^4$ order are $-\frac{3S'}{4M^4}p^2\frac{\textrm{d}}{\textrm{d}r}$, $-\frac{9S''}{16M^4}p^2$, and $\frac{3\Delta'}{16M^4}\frac{\kappa}{r}p^2$ shown in Table.~VIII in Ref.~\cite{Guo2019Phys.Rev.C99.054324}.
It is found that the sum of the contributions from these three terms are positive (negative) for the spin-up states with $\kappa<0$ (the spin-down states with $\kappa>0$), and the absolute value becomes bigger with the increase of $l$ (or $\kappa$).
Consequently, the large $l$ (or $\kappa$) states show the biggest discrepancy here, and the $\kappa<0$ ($\kappa>0$) states show negative (positive) discrepancy from the corresponding exact values here.
A possible way to improve the results further is to derive the $1/{M^\ast}^4$ order in the reconstituted FW transformation, and pay attention to the terms proportional to  $\frac{1}{{M^\ast}^4}p^2\frac{\textrm{d}}{\textrm{d}r}$, $\frac{1}{{M^\ast}^4}p^2$, and $\frac{1}{{M^\ast}^4}\frac{\kappa}{r}p^2$.

By summing up all the single-particle energies of the occupied states for $126$ neutrons in $^{208}$Pb, the exact value is $-3045.501$~MeV, while the result obtained by the $1/{M^\ast}^3$-order reconstituted FW transformation is $-3055.767$~MeV.
The corresponding relative discrepancy is only about $0.34\%$.

\subsection{Single-particle and total densities}\label{sec:IIIB}

\begin{figure}
  \includegraphics[width=0.45\textwidth]{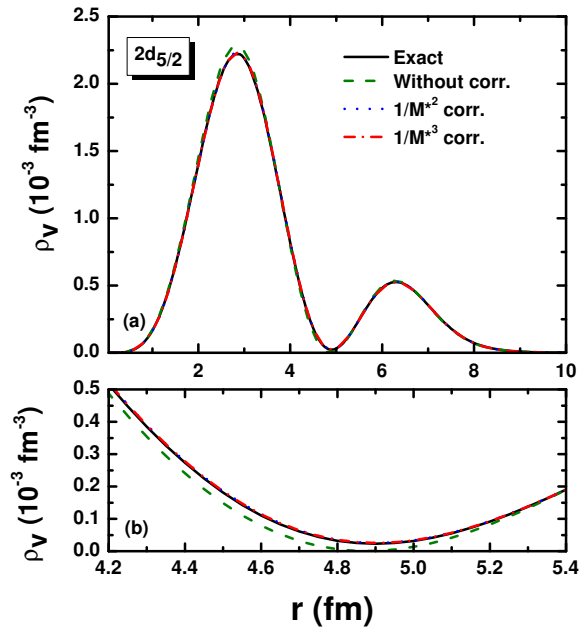}
  \caption{(Color online) (a) Single-particle density for the neutron $2d_{5/2}$ state.
  The solution of the Dirac equation~\eqref{eq:Dirac1} is shown with the black solid line, while the result by Eq.~\eqref{rhov1} is shown with the olive dashed line.
  The results by Eq.~\eqref{rhov2} with the corrections up to the $1/{M^\ast}^2$ and $1/{M^\ast}^3$ orders are shown with the blue dotted and red dash-dotted lines, respectively.
  (b) Details around the node of wave function in a smaller scale.}\label{fig2}
\end{figure}

For the discussion on the single-particle density, let us first take the neutron $2d_{5/2}$ state as an example, which has one-node structure and is neither deeply nor weakly bound.
The exact single-particle density for the $2d_{5/2}$ state is shown with the black solid line in Fig.~\ref{fig2}.
For comparison, the single-particle density calculated by Eq.~\eqref{rhov1} without any relativistic correction (i.e., picture-change-error correction) is shown with the olive dashed line.
In contrast, the results calculated by Eq.~\eqref{rhov2} with the relativistic corrections up to the $1/{M^\ast}^2$ and $1/{M^\ast}^3$ orders are shown with the blue dotted and red dash-dotted lines, respectively.

For the result without correction, on the one hand, a deviation from the exact density is visible around the first peak of the wave function.
On the other hand, it is even more important to point out its deviation from the exact density around the node of the wave function, which is shown in Panel (b) of Fig.~\ref{fig2} in a smaller scale.
Because of the existence of the small component of Dirac spinor, i.e., $F(r)$ in Eq.~\eqref{eq:Dirac1}, the single-particle density in the Dirac picture, $\rho_v(r) = [G^2(r)+F^2(r)]/(4\pi r^2)$, will never become exactly zero at any finite radius $r$.
However, the corresponding single-particle density in the Schr\"odinger picture (Eq.~\eqref{rhov1}) becomes exactly zero at the nodes of the single-particle wave function $\varphi(r)$.
This is one of the explicit reasons why the relativistic corrections to the densities are indispensable.
In other words, the picture-change error belongs to the systematic error, instead of the matter of truncation or numerical accuracy.

By taking into account the relativistic corrections, the results obtained by Eq.~\eqref{rhov2} are almost identical to the exact density in the whole region, including the node region as shown in Panel (b) of Fig.~\ref{fig2}.
Indeed, in Eq.~\eqref{rhov2}, the terms such as
$$\varphi^\dag(r) \overleftarrow{\frac{\textrm{d}}{\textrm{d}r}} \frac{1}{4{M^\ast}^2} \overrightarrow{\frac{\textrm{d}}
{\textrm{d}r}} \varphi(r)$$
prevent the single-particle density from becoming exactly zero at any finite radius $r$.

\begin{figure}
  \includegraphics[width=0.45\textwidth]{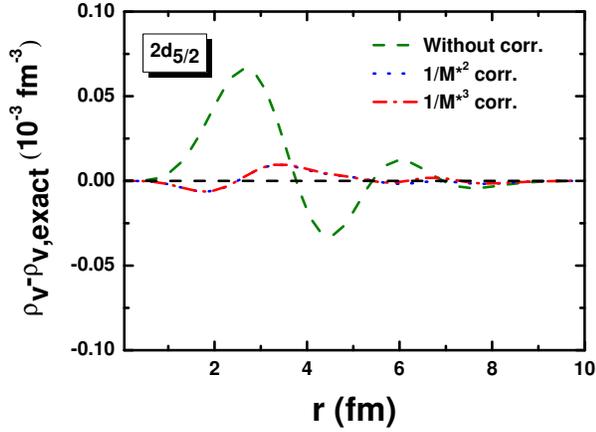}
  \caption{(Color online) Difference between the single-particle density of the $2d_{5/2}$ state calculated by the reconstituted FW transformation and the corresponding exact density.
  The choice of legends follows that in Fig.~\ref{fig2}.
 }\label{fig3}
\end{figure}

In order to examine the details, the differences between the single-particle density of the $2d_{5/2}$ state calculated by the reconstituted FW transformation and the corresponding exact density are shown in Fig.~\ref{fig3}.
It can be seen clearly that the results without the corrections show dramatic differences from the exact ones, and the discrepancy $\rho_v - \rho_{v,{\rm exact}}$ reaches $6.6 \times 10^{-5}$~fm$^{-3}$ around the first peak of the wave function, which corresponds to a relative discrepancy $3.2\%$.
In contrast, after considering the picture-change-error corrections, the discrepancies $\rho_v - \rho_{v,{\rm exact}}$ are smaller than $0.97 \times 10^{-5}$~fm$^{-3}$ in the whole region.

\begin{figure}
  \includegraphics[width=0.45\textwidth]{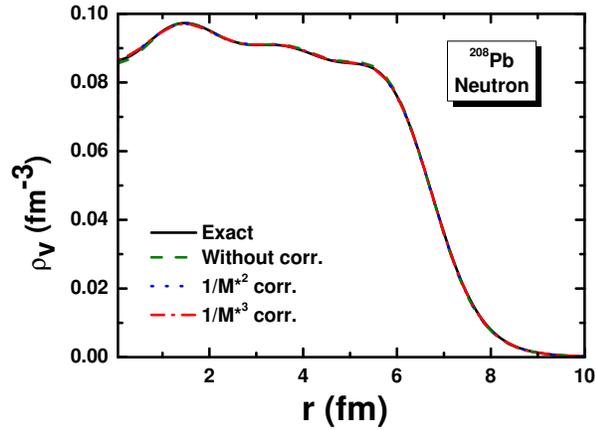}
  \caption{(Color online) Neutron density of $^{208}$Pb calculated by the reconstituted FW transformation and the corresponding exact density.
  }\label{fig4}
\end{figure}

Since the non-relativistic expansion is regarded as a crucial candidate for the connection between the relativistic and non-relativistic DFT, it is meaningful to further examine the total density of neutron by summing up the single-particle densities of the lowest $126$ single-particle orbitals as shown in Fig.~\ref{fig4}.
In the center area with $r<6$~fm, the total neutron density lies between $0.08$~fm$^{-3}$ and $0.10$~fm$^{-3}$.
As seen in the figure, all the results calculated by the reconstituted FW transformation are in good agreements with the exact result.

\begin{figure}
  \includegraphics[width=0.45\textwidth]{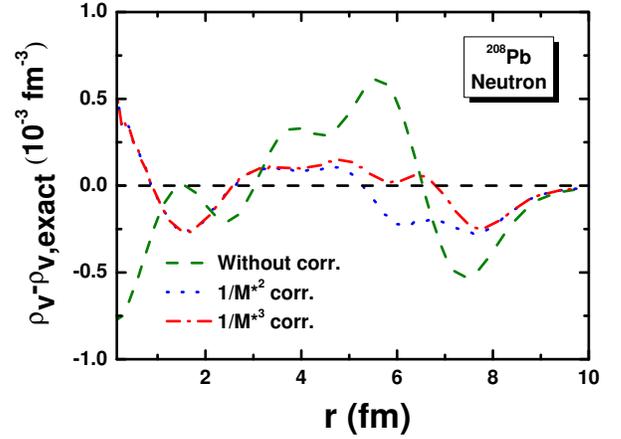}
  \caption{(Color online) Difference between the neutron density of $^{208}$Pb calculated by the reconstituted FW transformation and the corresponding exact density.}\label{fig5}
\end{figure}

In order to investigate the results in Fig.~\ref{fig4} in detail, the differences between each calculated result and the exact neutron density are shown in Fig.~\ref{fig5}.
It can be seen that the results calculated with the consideration of the relativistic corrections show quite different behaviors from the one without corrections.
In general, the relative discrepancy between the results with corrections and the exact one is within $0.5\%$.
It is also interesting to point out that, compared with the result with the corrections up to the $1/{M^\ast}^2$ order, the neutron density is further improved by the $1/{M^\ast}^3$-order corrections in the nuclear surface region, since the $1/{M^\ast}^3$-order corrections are proportional to the derivatives of the potentials as shown in Eq.~\eqref{rhov2}.

\subsection{Single-particle and total scalar densities}\label{sec:IIIC}

\begin{figure}
  \includegraphics[width=0.45\textwidth]{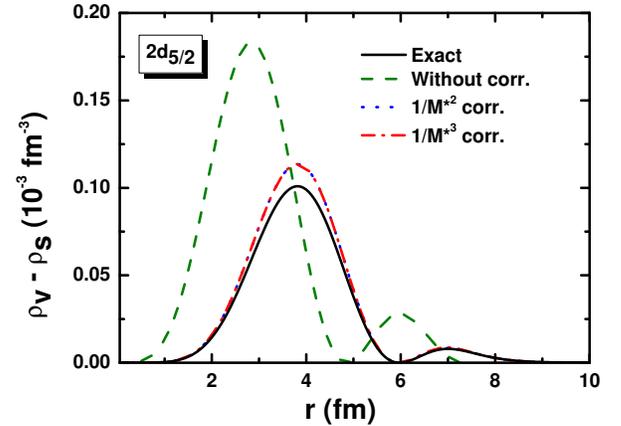}
  \caption{(Color online) Difference between the single-particle density and the single-particle scalar density for the neutron $2d_{5/2}$ state.
   The solution of the Dirac equation~\eqref{eq:Dirac1} is shown with the black solid line, while the result by Eq.~\eqref{rhos1} is shown with the olive dashed line.
  The results by Eq.~\eqref{rhos2} with the corrections up to the $1/{M^\ast}^2$ and $1/{M^\ast}^3$ orders are shown with the blue dotted and red dash-dotted lines, respectively.}\label{fig6}
\end{figure}

For the discussions on the single-particle scalar density, let us first focus on the difference between the single-particle density and the single-particle scalar density.
In the Dirac picture, such a difference comes from the small component of Dirac spinor, i.e., $\rho_v(r) - \rho_s(r) = 2F^2(r)/(4\pi r^2)$ in the present case, which is in general tiny.
This tiny difference is shown with the black solid line in Fig.~\ref{fig6}, by taking the $2d_{5/2}$ state as an example.
The peak of the black solid line is only around $0.10 \times 10^{-3}$~fm$^{-3}$ at $r=3.75$~fm.

For comparison, the corresponding result by Eq.~\eqref{rhos1} without any relativistic (picture-change-error correction) correction is shown with the olive dashed line in Fig.~\ref{fig6}.
In contrast, the results calculated by Eq.~\eqref{rhos2} with the relativistic corrections up to the $1/{M^\ast}^2$ and $1/{M^\ast}^3$ orders are shown with the blue dotted and red dash-dotted lines, respectively.

\begin{figure}
  \includegraphics[width=0.45\textwidth]{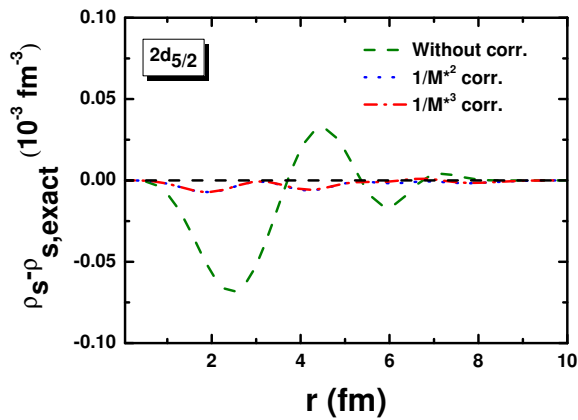}
  \caption{(Color online) Difference between the single-particle scalar density of the $2d_{5/2}$ state calculated by the reconstituted FW transformation and the corresponding exact density.
  The choice of legends follows that in Fig.~\ref{fig6}.}\label{fig7}
\end{figure}

It can be seen in Fig.~\ref{fig6} that the differences between the results without the relativistic corrections and the exact one are systematic.
There are differences in the positions of peaks and nodes as well as the amplitudes.
In contrast, the results with the relativistic corrections reproduce remarkably the behavior of the exact result, in particular the positions of peaks and nodes.
The remaining discrepancies are rather small (notice the small scale used here).

As a result, the exact single-particle scalar density for the $2d_{5/2}$ state can be well reproduced, with a relative discrepancy less than $0.2\%$, by the reconstituted FW transformation including the relativistic corrections, as illustrated in Fig.~\ref{fig7}.

Comparing the dashed lines to the others in Figs.~\ref{fig6} and \ref{fig7}, one can confirm the critical role of the relativistic corrections playing in the difference between the vector and scalar densities, because they both start from the $1/{M^\ast}^2$ order.

\begin{figure}
  \includegraphics[width=0.45\textwidth]{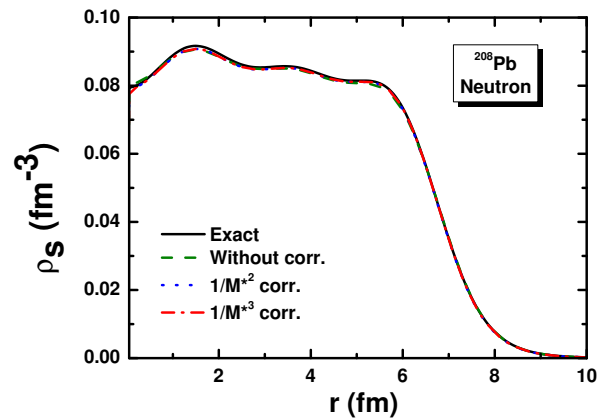}
  \caption{(Color online) Neutron scalar density of $^{208}$Pb calculated by the reconstituted FW transformation and the corresponding exact density.}\label{fig8}
\end{figure}

\begin{figure}
  \includegraphics[width=0.45\textwidth]{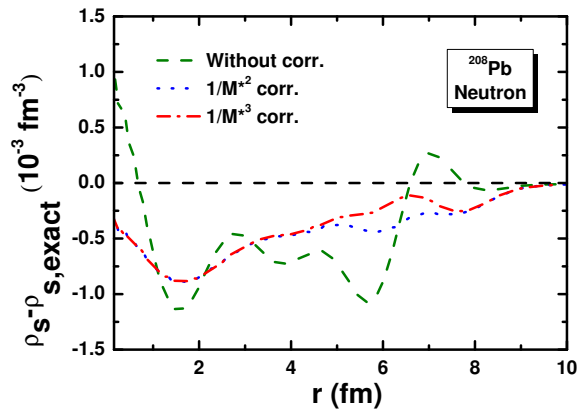}
  \caption{(Color online) Difference between the neutron scalar density of $^{208}$Pb calculated by the reconstituted FW transformation and the corresponding exact density.}\label{fig9}
\end{figure}

The total neutron scalar density of $^{208}$Pb is shown in Fig.~\ref{fig8}.
It is smaller than the total neutron density shown in Fig.~\ref{fig4} by $3\sim8\%$ in general.
For the details, the differences between the neutron scalar densities calculated by the reconstituted FW transformation and the exact result are shown in Fig.~\ref{fig9}.
The results with the relativistic corrections can well reproduce the exact values, with a relative discrepancy less than $1\%$.
The same as the total density, the $1/{M^\ast}^3$-order corrections can further improve the agreement with the exact result in the nuclear surface region.

\section{Summary and Perspectives}\label{sec:IV}

With the inspiration of the reconstituted SRG method \cite{Guo2019Phys.Rev.C99.054324}, the reconstituted FW transformation has been proposed.

In the covariant DFT, since the inclusion of the strong scalar potentials, the conventional FW transformation shows a quite slow convergence for the spectrum of the single-particle energies.
By replacing the bare mass $M$ with the Dirac mass $M^\ast$ and defining the corresponding new operators, the contributions that come from the products of the terms that already appear in the lower orders and $\left(\frac{S}{M}\right)^n$ ($n=1, 2, \cdots$) are absorbed into the lower orders.
As a result, the convergence of the spectrum obtained by the reconstituted FW transformation becomes much faster than the conventional one.
In addition, since the reconstituted FW transformation starts from the general operators, it provides the foundation of the reconstituted SRG method and leads to a more promising future of application.

Besides the single-particle energies, the single-particle and total vector and scalar densities by the reconstituted FW transformation also reproduce the exact result well.
In particular, the relativistic corrections to the densities have been investigated in detail.
The corrections actually come from the picture-change error between the Schr\"odinger and Dirac pictures.
The corrections to the single-particle densities start appearing from the $1/{M^\ast}^2$ order, and they prevent the single-particle density from becoming exactly zero at any finite radius r.
Moreover, the derivations between $\rho_{v,\textrm{non}}$ and $\rho_{s,\textrm{non}}$ appears from the $1/{M^\ast}^2$ order, and the relativistic correction $\Delta \rho_s$ also appears from the same order.
Therefore, the relativistic corrections play a crucial role in the difference between the vector and scalar densities.
The results thus obtained with the consideration of the corrections are almost identical to the exact results.

Based on the above discussions, the reconstituted FW transformation paves a promising way for the connection between the relativistic and non-relativistic density functional theories for the future studies.

\begin{acknowledgments}

The authors are grateful to Mr.~Tomoya Naito, Dr.~Zhengxue Ren, and Professor Pengwei Zhao for the helpful discussions on the relativistic corrections to the single-particle densities.
This work was partially supported by the JSPS Grant-in-Aid for Early-Career Scientists under Grant No.~18K13549, the JSPS-NSFC Bilateral Program for Joint Research Project on Nuclear mass and life for unravelling mysteries of the $r$-process, and RIKEN Pioneering Project: Evolution of Matter in the Universe.
Y.G. also acknowledges the scholarship from Asian Future Leaders Scholarship Program funded by Bai Xian Asia Institute.

\end{acknowledgments}

\end{document}